\documentclass[letterpaper]{JHEP}
\usepackage{epsfig}
\usepackage[active]{srcltx}

\newcommand{\be}{\begin{equation}}
\newcommand{\ee}{\end{equation}}
\newcommand{\bea}{\begin{eqnarray}}
\newcommand{\eea}{\end{eqnarray}}
\newcommand{\half}{{1 \over 2}}

\def\IZ{\relax\ifmmode\mathchoice
{\hbox{\cmss Z\kern-.4em Z}}{\hbox{\cmss Z\kern-.4em Z}}
{\lower.9pt\hbox{\cmsss Z\kern-.4em Z}}
{\lower1.2pt\hbox{\cmsss Z\kern-.4em Z}}\else{\cmss Z\kern-.4em
Z}\fi}
\def\IR{\relax{\rm I\kern-.18em R}}
\font\cmss=cmss10 \font\cmsss=cmss10 at 7pt

\newcommand{\pq}{$(p,q)$ }

\preprint{TAUP-2539-98\\{\tt hep-th/9812021}}

\title{Thermal Monopoles}

\author{Barak Kol
\\
School of Physics and Astronomy\\ Tel Aviv University\\ Ramat Aviv 69978,
Israel\\
\email{barak1k@post.tau.ac.il}
}

\abstract{Evidence is given that some particles with large
charges in gauge theories, including 4d ${\cal
N}=4~~SU(3)$ and 5d theories with ${\cal N}=1$
supersymmetry, have macroscopic entropy. The ground state
entropy is found to be extensive in the charges. The mass
gap for excitation is estimated and is of the same order
as the binding energy. The question of the decay of
excited states through Hawking radiation is raised. The
analysis is based on the model of
\pq string webs and on various other results. The evidence
points to an explanation in terms of a string model.}

\begin{document}

\section{Introduction}

Consider the degeneracy of BPS multiplets in 4d $N=4$ and
5d $N=1$ gauge theories, which can both be represented by
\pq  string webs \cite{AHK,KolRahmfeld,BergmanKol}. In 4d,
the simplest example is a spontaneously broken gauge group
$SU(3) \to U(1)^2$, and in 5d one can consider a
spontaneously broken $SU(2)$. Since such particles carry
both electric and magnetic charges (in 4d), we may refer
to them generically as `monopoles'. In some cases, I find
evidence that their degeneracy is exponential in the
charges, $q_i$ and (to leading order) the exponent is
extensive, that is, homogeneous of degree one. Thus the
entropy of the ground state satisfies
\be
S(\lambda q_i)=\lambda S(q_i), \; \forall
 \lambda \in \IR^+
\label{entropy}
\ee
(see note added at the end of the introduction regarding
the constant of proportionality).

After a review of string webs in section \ref{review}, the
evidence is detailed in section \ref{evidence}. The direct
evidence is for the 1/2 BPS states in 5d, and the
conjectured extension to 1/4 BPS states in 4d relies on
their common web model. First I present two experimental
results which imply the proposed entropy formula. A. Klemm
\cite{KlemmPriv,Chiang:1999} computed the sizes of some
multiplets in 5d from the Seiberg-Witten curve
\cite{SW,KKV9609,BK5vM}, and their asymptotic growth was
later determined in \cite{KlemmZaslow}. The asymptotic
growth of a 4d ${\cal N}=2$ spectrum was found in
\cite{MNW9707} in a model with origins in tensionless
strings \cite{GanorHanany,KMV9607}. The second
argument\footnote{Due to J. Maldacena.} is an analogy with
an 8 dimensional black hole, which is a string web on an
internal torus \cite{SenNet}. This black hole has a
perturbative dual (with winding and momentum charges along
a circle) and thus its entropy is known exactly. For large
charges one would expect the entropy to agree with the
monopole's, as they are both described by large webs. The
third argument uses 6d strings \cite{GanorHanany} and
their suggested role in the BPS particle spectrum
\cite{W_phase} to estimate the entropy. {\it Note that
almost all methods assume a string model to explain the
entropy}. All methods agree with (\ref{entropy}).

Having entropy, these particles seem to be thermal
systems. In contrast, the BPS spectrum of other gauge
theories is known not to contain any large degeneracies.
In 4d $N=4$ with an $SU(2)$ gauge group, the BPS spectrum
consists of the
\pq dyons which all have the same multiplet size as they are all
$SL(2,Z)$ dual. In 4d $N=2$ Seiberg-Witten theories large
multiplets are not known either. For instance, in pure
$SU(2)$ the BPS spectrum consists of W particles in a
vector-multiplet, and dyons in a hyper
\cite{SW,BilalFerrari9602}. In theories with gravity, on
the other hand, black holes which are thermal bodies are
the common case, as was realized in the 70's by Bekenstein
and Hawking \cite{Bekenstein1,Bekenstein2,HawkingTemp}.

In section \ref{MassGap} the web picture is used to estimate the mass gap
\be
\Delta M \sim M_W/q,
\ee
where $M_W$ is the mass of some (perturbative) particle
such as the W, and $q$ is a typical charge. It is known
that the 5d states are only marginally stable, whereas the
4d states are stable in general. However, in the 4d case,
the mass gap is found to be of the same order as the
binding energy, and thus as far as the excited spectrum is
concerned {\it the 4d states are effectively marginally
stable as well}. In comparison, in gravity also extremal
black holes are energetically only marginally stable, but
they are at least ``entropically stable" since a fission
of an extremal black hole with a non-zero horizon, results
in an increase in the horizon area. {\it These monopoles,
however, are not even entropically stable because of
(\ref{entropy})}.

I would like to raise a few questions regarding the absorption and
emission of radiation by these monopoles.  Using the
analogy with black holes one finds an expression for the entropy as a
function of a small excitation energy $\epsilon \ll M$ (section \ref{BH})
\be
S(\epsilon,q_i) \simeq \sqrt{\epsilon/\Delta M} + S_0(q_i)
\ee
where $S_0$ is the ground state entropy, and $\Delta M$ is the mass
gap. Based on this relation the temperature is
\be
T \simeq \sqrt{\epsilon \cdot \Delta M}
\ee
For the extremal state the temperature is zero, of course.
The analogy {\it seems to predict Hawking evaporation of excited
  monopoles}, and that the cross section would be proportional to the
entropy. However, as the ground state is only marginally
stable (even entropically), the excited state may prefer
to fission rather than evaporate (see \cite{DasMathur} for
a study of the same question in the case of black holes
which concluded that there evaporation dominates).

In order to go beyond the black hole analogy one needs a
thermodynamic description for the monopoles.
In gravity, black holes were discovered to be thermal bodies in the
macroscopic framework of general relativity in
the 70's, while only in '96 was some microscopic description found, using
string theory \cite{StromingerVafa,CallanMaldacena} (and many others,
for a recent review see \cite{PeetRev}). Here the case seems to be
reversed - the microscopic string description is used to motivate
research in field theory.

The thermodynamics may be constructed either in the
microscopic string picture or in field theory. Working in
field theory, however, one encounters an immediate problem
as general classical solutions for the monopoles are not
available. Some 4d solutions were found \cite{HHS,LeeYi},
but solutions for large charges are still unknown. In
particular, not much is known about their size and shape
in space-time, and recently this problem was studied in
\cite{KolKroyter}. Another obstacle on the way to a
semi-classical description is that the coupling is often
of order one, otherwise most of the states are outside
their domain of stability\footnote{Recall that the
monopoles of 4d $N=4$ exist only within a certain wall of
marginal stability\cite{BergmanWeb}, analogous to the
curve of marginal stability in 4d  $N=2$.}. In 5d the
situation is even worse as there is no knowledge of the
classical solutions, and the Yang-Mills is
non-renormalizable.

Note added: since the appearance of the first version of
this paper on the hep-th archive, a number of relevant
papers appeared. In \cite{Chiang:1999,KlemmZaslow} the
asymptotic growth of the degeneracy was analytically found
for some 5d models. These results confirm the proposed
macroscopic entropy (eq. \ref{entropy}). However, the
constant of proportionality is different than the one
which was phenomenologically suggested in the first
version
\be
S=2 \pi \sqrt{2 F(q_i)}
\label{entropy2}
\ee
where $F(q_i)$ is the number of faces in the web, a
quadratic function in the charges, which can be explicitly
determined by a grid diagram \cite{AHK}.  Even though the
proportionality constant remains to be fixed this is
unimportant to the basic picture suggested here.

More recently, a Witten index calculation for the
degeneracy of 4d 1/4 BPS states \cite{SternYi} confirmed
that there is a large degeneracy of irreducible SUSY
representations. However, since the results are not
applicable to the case of taking all charges to be large,
a quantitative study of the entropy was not possible.

\section{Review of String Webs}
\label{review}

\begin{figure}
\centerline{\epsfxsize=80mm\epsfbox{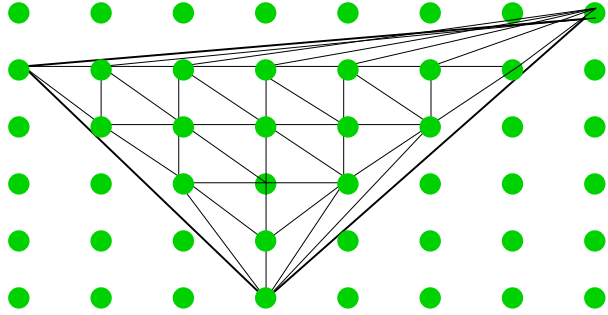}}
\medskip
\caption{The 4d example: a grid diagram for a monopole with charges
 $(-1,7) \; (-4,-3) \; (5,-4)$ in  $SU(3) \to U(1)^2$. This web has
 $F=15$ faces.}
\label{fig4d}
\end{figure}

Let us consider the simplest configurations of particles described by
string webs in 4d and 5d. In 4d, consider an $N=4$ theory with a
spontaneously broken gauge group $SU(3) \to U(1)^2$. This theory has a
brane realization as 3 D3 branes in type IIB, in the
field theory limit
\be
\Delta x/l_s^2=\Phi=\textnormal{const},\; M_s=1/l_s \to \infty,
\ee
where $\Delta x$ are distances in the $\IR^6$ transverse
to the brane, and $l_s$ is the string length\footnote{We
do not  have to consider the decoupled center of mass
$U(1)$ factor.}. The moduli space of the theory is given
by the locations of the D3 branes (which are also the vevs
of the adjoint Higgs), namely 3 six-vectors $\Phi_i$  (up
to permutations). A string web ending on the D3s describes
a $1/4$ BPS particle in the gauge theory (for a definition
of a string webs see \cite{BergmanKol} or \cite{AHK}). Its
BPS mass is given by summing the mass of the individual
edges \cite{BergmanWeb}. The charges of the particle are
given by the \pq labels of the external strings: a \pq
string ending on a D3 carries $p$ units of electric charge
and $q$ units of magnetic charge under the $U(1)$ factor
associated with that brane. Although we seemingly defined
3 sets of charges, there is one constraint, that the sum
of both electric and magnetic charges vanishes. A grid
diagram for a typical large web\footnote{Recall that the
grid diagram is the dual diagram to the web, exchanging
vertices with faces.} with charges $(-1,7) \; (-4,-3) \;
(5,-4)$ and $F=15$ faces is shown in figure \ref{fig4d}.

\begin{figure}
\centerline{\epsfxsize=100mm\epsfbox{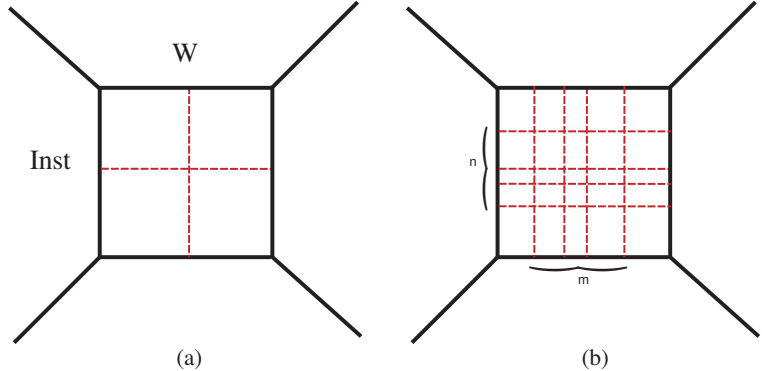}}
\medskip
\caption{The 5d example: $SU(2) \to
U(1)$. (a) The W and the instanton. (b) A bound state of
 $m$ W's and $n$ instantons.}
\label{fig5d}
\end{figure}

In 5d, consider an $N=1$ theory with a spontaneously broken $SU(2) \to
U(1)$ (and no $\Theta$ angle). This theory can be realized by a
5-brane web (in the field theory limit), shown as thick lines in
figure \ref{fig5d}. In this picture, the W is represented by a vertical
fundamental string, and the instanton (which is a particle in 5d) is
represented by a vertical D-string. We would like to consider a
(marginally) bound state of $m$ W's and $n$ instantons, represented
by a string web ending on the 5-brane web, shown as dashed lines in
figure \ref{fig5d}.

Given a web one can read off the number of faces, $F$. Actually, it is
enough to know the charges of the external strings and to use grid
diagrams \cite{AHK}. In the 5d example we have $F=(n-1)(m-1)$, while
in the 4d case we have
\be
F={1 \over 6} \det \left[ \matrix{p_1& q_1& 1\cr p_2& q_2& 1\cr p_3& q_3& 1} \right]+1-n_X/2
\ee
where $n_X$ is the number of external strings, usually 3 in our example.

The web representation is known to predict the (maximal) spin of the
particle \cite{KolRahmfeld,BergmanKol}. In order to find the multiplet
structure one needs to find the bosonic and
fermionic zero modes of the web, and solve for the ground states of
the supersymmetric quantum mechanics on the moduli space of the web. Space-time
spin is carried only by the fermionic zero modes (FZM), and their
number is found to be
\be
n_{FZM}=8F+4n_X
\ee
Each pair of FZM produces a fermionic oscillator which may raise the
spin by $\half$, so one finds the maximal spin to be
\be
j \le n_{FZM}/8 = F + n_X/2
\label{max_spin}
\ee

Even though we know the bosonic and fermionic coordinates
on the moduli space of the web, we do not know in general
to find the ground states, that is, the full multiplet
structure. The complication in this approach arises
because the moduli space has a boundary and because some
fermions do not have bosonic partners. Nevertheless, in
the next section I will provide evidence for the claimed
asymptotic growth of states.

For some special cases the multiplet structure is known.
For 5d $F=0$ it is an irreducible SUSY representation
constructed by tensoring  an $SU(2)_L$ representation with
spin $j_{max}-1/2$ (\ref{max_spin})
 with a half hyper $SU(2)_R$ BPS representation \cite{W_phase,KolRahmfeld}.
For 4d $F=0$ (without coinciding external legs) one finds
a representation of size $2^{n_{FZM}/2}$ from $n_{FZM}/2$
free fermionic oscillators, a representation which
includes the medium SUSY representation as a factor
\cite{BergmanKol}. For non-zero $F$, the problem was
solved for a special case, the 5d example with $n=2$
instanton number and arbitrary electric charge $m$, by
Nekrasov and the author \cite{NekrasovWoods}, by using
additional bosonic Wilson loop coordinates suggested by
Vafa \cite{VafaPriv,GopakumarVafa}. One finds that the
representation is a tensor product of an $SU(2)_L$
representation (with degeneracy $2l+1=2(n+m)$), the BPS
half hyper, and a third multiplet given by the level $m-1$
of 5 bosonic oscillators, 4 with ``base frequency" and one
with a double frequency. The degeneracy (after factoring
out the BPS half-hyper) is
\be
d_{m,2}=2(m+2) {1 \over (1-q)^4 (1-q^2)} {\biggr
\vert}_{q^{m-1}}. \; \; \; \footnote{The expected spin content, after a reduction to 4d, is that the 4
 base oscillators are in a triplet and a singlet, while the double
 oscillator is in a singlet. Namely, in the partition function we should replace
  $(1-q)^4 (1-q^2) \to (1-s^2q)(1-s^{-2}q)(1-q)^2(1-q^2)$, where the coefficient of
  $s^{2j_z}$ in the expansion counts the number of states having that value of
  $j_z$.}
\label{degm2}
\ee
This is a phenomenological formula that fits the
experimental data of
 \cite{KMV9607}.
It should have an interpretation as creating the $m$
W's perturbatively in the 2 instanton background.

Note that formula (\ref{degm2}) as well as the expressions
in \cite{KMV9607} give polynomial growth with $m$, but
there is no contradiction with (\ref{entropy}), since the
latter requires both $m$ and $n$ to be large. An example
for a function with this property is $(m+n)!/(m! ~n!)$.

\section{Evidence for Entropy}
\label{evidence}
\subsection{Experimental data}
The degeneracies of multiplets\footnote{Actually, their
  Witten index.} can be computed from the Seiberg-Witten curve, as they
are  encoded in the expansions of the periods. For the
$E_0$ 5d theory, A. Klemm finds the degeneracies
\cite{KlemmPriv,Chiang:1999}
\be
\matrix{\underline{d_{1-5}}& \underline{d_{6-10}}&
\underline{d_{11-15}}& \underline{d_{16-20}} \cr
3& -17062& 4827935937& -2927443754647296 \cr
-6& 188464& -66537713520& 44000514720961743 \cr
27& -2228160& 938273463465& -668908727886779298 \cr
-192& 27748899& -13491638200194& 10272581487272296287 \cr
1695& -360012150& 197287568723655& -159199764298612184400
}
\ee
where the subscript of $d$ is the charge under the (only)
$U(1)$ gauge group in the theory. The first column titled
$d_{1-5}$ specifies $d_1$ to $d_5$ from top to bottom, and
similarly for the other columns. This is equivalent to
counting curves on ${\bf P}^2$.

\begin{figure}
\centerline{\epsfxsize=100mm\epsfbox{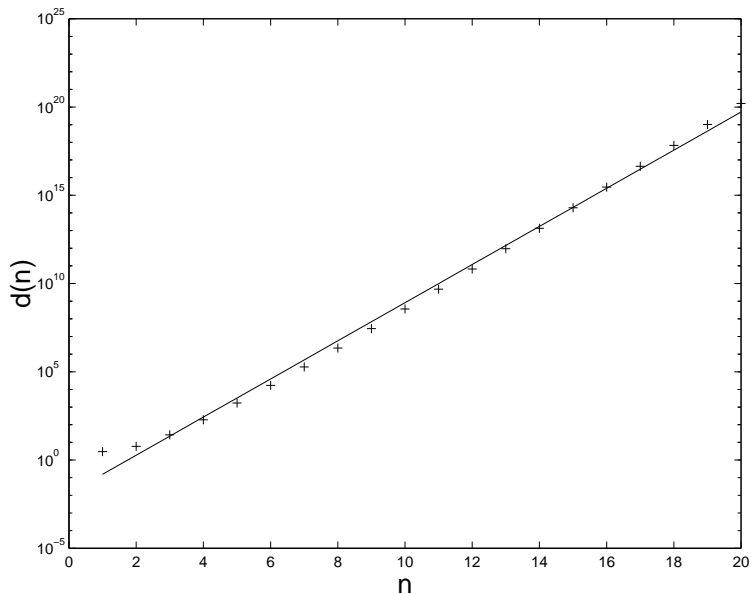}}
\medskip
\caption{A semi-log plot of the degeneracies $d(n)$ of
 BPS multiplets in the 5d E0 theory for charges $n=1...20$.
The +'s are data and the line is a linear fit. We note
that the entropy $S=\log(d)$ is linear in $d$.}
\label{E0exp}
\end{figure}

Graphing $\log(d_n)$ as a function of $n$ one gets a
straight line, which is evidence to the claim
(\ref{entropy}) - see figure \ref{E0exp}. The slope of the
fitted line is 2.49. An analytic argument
\cite{KlemmZaslow} yields $2\pi*0.4627 \simeq 2.68$ for
the asymptotic slope, which is in reasonable agreement.

Another computation \cite{MNW9707}, this time in 4d, takes
a 6d tensionless string theory described by F theory on an
elliptically fibered ${\bf F}_1$, with the ${\bf F}_1$
shrinking, compactified to 4d on a torus. Here ${\bf F}_1$
stands for the first Hirzebruch surface, a non-trivial
fibration of ${\bf P}^1$ over ${\bf P}^1$. BPS states are
characterized by two charges $n_1,n_2$ and the authors
find the asymptotic degeneracy to be
\be
\log (d(n_1,n_2))=2 \pi \sqrt{n_1^2+2 n_1 n_2} - {7 \over 4}\log (n_1^2+2 n_1 n_2).
\ee
This expression agrees with (\ref{entropy}). To extract
the constant of proportionality recall that for ${\bf
F}_1$ the number of faces is $2F=n_1^2+2 n_1 n_2$
(\cite{KolRahmfeld}, for example), and so we get agreement
with (\ref{entropy2}).

\subsection{Black hole analogy}
\label{BH}
Consider a string web on a torus, or a periodic web, as an
approximate model for a monopole described by a large web.
In case the number of boundary edges is small compared to
the number of faces in a web, one expects this to be a
good approximation.

Specifically, consider a web on a right angled torus, with
$n$ D-strings winding horizontally and $m$ strings winding
vertically (see figure \ref{per_web}a). It is a D-brane
model for an 8d extremal black hole (with no classical
area). The number of faces is $F=mn$. We can use dualities
to get a perturbative system - T dualizing along both
directions and then further S dualizing we get a
fundamental string wound $n$ times and carrying $m$ units
of momentum. The entropy is
\be
S=2 \pi \sqrt{{c_R \over 6}m}=2 \pi \sqrt{2mn}=2 \pi \sqrt{2F}
\ee
 in agreement with (\ref{entropy}) with the constants as in (\ref{entropy2}).

The mass gap corresponds to exciting one unit of left
moving fractional momentum and one right moving and is
given by $\Delta M=2M/(m n)$ (see (\ref{gap}). If we
excite the string by adding an energy $\epsilon$ with
$\delta=\epsilon /\Delta M$ units of both left and right
moving momentum, the entropy is
\be
S=2 \pi \sqrt{2mn+2\delta}+ 2 \pi \sqrt{2 \delta}.
\ee

\subsection{6d strings}
In 5d the angular momentum, $J$ of the BPS spectrum scales
with the mass as
\be
J \sim M^2.
\ee
In terms of webs this is understood from $J \sim F, ~M
\sim \sqrt{F}$. In \cite{W_phase} it was observed that this is the
same relation as for a string, and the spectrum was
suggested to have a 6d origin in terms of a 6d string
which becomes tensionless at the conformal point. If we
try the string model to estimate the entropy we get
\be
S \sim M \sim \sqrt{F}
\ee
in agreement with (\ref{entropy}).

\section{Mass Gap}
\label{MassGap}
\begin{figure}
\centerline{\epsfxsize=100mm\epsfbox{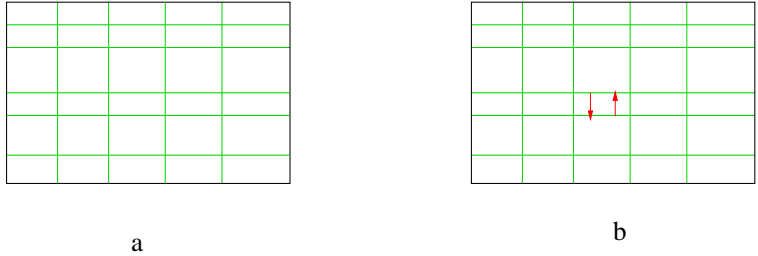}}
\medskip
\caption{ (a) A periodic BPS web. (b) The lowest excitation. The pair
of net charge zero is denoted by arrows.}
\label{per_web}
\end{figure}

Let us estimate the mass gap, or the excitation energy step, for the
5d example, which is expected to be typical.
Looking at the string web one identifies the lowest excitation to be a
pair creation of two opposite string edges in the web (see figure
\ref{per_web}b). Actually there are two options to choose from -
exciting either
vertical or horizontal strings. Suppose a fractional vertical string
is lighter, then the excitation energy is roughly twice the full length
string, $M_W$ (or some basic particle mass in general)
divided by the number of vertical strings, $n$ (or some other charge
$q$ in general)
\be
\Delta M= {2 M_W \over n}= ~{2M \over m n}.
\label{gap}
\ee
The same result can be reached by considering a periodic
web, where by dualities the system is mapped to a wound
string. There the smallest excitation is known to be a
pair of left and right moving fractional momentum
(assuming a unit of fractional momentum is lighter than a
unit of fractional winding), which gives the same estimate
for the mass gap.

Let us compare the mass gap with a typical binding energy in 4d. The
estimate we find is
\be
M_{bind} \sim {M \over q^2}
\label{binding}
\ee
where $q$ is a typical integral charge. Thus the binding
energy is of the same order as the energy gap (\ref{gap})
so there are at most a few excited bound states.

To derive ({\ref{binding}) consider the BPS mass formula
$M^2=Q_e^2 + Q_m^2 +2
\left| Q_e
\times Q_m \right|$, where $Q_e,\,Q_m$ are the electric
and magnetic charge two-vectors given by
 $\vec{Q}_e+i\vec{Q}_m=\sum_{i=1}^{3}{(p_i+\tau q_i)
\vec{\Phi}_i}$ and we will consider order 1 coupling, $\tau \simeq i$.
 Given a state with some
charges \pq, consider its binding energy with respect to a
decay into two states with roughly half the charge. Since
the charges may be odd we should consider a decay into
particles A and B of charges $(p,q)_A=(p,q)/2+\delta,\;
(p,q)_B=(p,q)/2-\delta$, where $\delta$ are order 1
adjustments. Now compute the binding energy
$M_{bind}=M_A+M_B-M$. The $O(\delta^0)$ term vanishes
because $M$ is extensive. The first order in $\delta$
vanishes because the expression is even in $\delta$. So
$M_{bind} \sim
\delta^2 \cdot d^2 M/d{\delta}^2$. To estimate the second derivative assume
$M \sim Q_e \sim Q_m$ (valid for order one coupling), and $dQ_{e,m}/d
\delta \sim Q_{e,m}/q$ so that each derivative contributes a $1/q$
factor, thus getting (\ref{binding}).

\vspace{.7cm}
\centerline{\Large {\bf Acknowledgements}}
\vspace{.4cm}

I am grateful to Albrecht Klemm, Juan Maldacena, Nikita
Nekrasov and Cumrun Vafa for valuable discussions. I would
like to thank several people at different institutions for
discussions and hospitality: E. Gimon, K. Hori, D. Kabat,
M. Rozali, S. Sethi P. Yi, other participants and the
organizers of the Aspen workshop; A. Hanany and W. Taylor
at MIT; A. Lawrence at Harvard; O. Aharony and A.
Rajaraman at Rutgers ; and finally I thank Boaz Kol, Y.
Sonnenschein and S. Yankielowicz at TAU for their help.

Work supported in part by the US-Israel Binational Science
 Foundation,
by GIF - the German-Israeli Foundation for Scientific Research,
and by the Israel Science Foundation.

\bibliography{thermal}
\bibliographystyle{JHEP}

\end{document}